\documentclass{article}
\usepackage{cite}
\usepackage{graphicx}
\usepackage{dcolumn}

\begin{document}

\date{}
\title{On the spectrum of the screened Coulomb potential $V(r)=-r^{-1}e^{-C/r}$}
\author{Francisco M. Fern\'{a}ndez\thanks{%
fernande@quimica.unlp.edu.ar} \\
INIFTA, DQT, Sucursal 4, C. C. 16, \\
1900 La Plata, Argentina}
\maketitle

\begin{abstract}
We analyse recent contradictory results and conclusions about the spectrum
of the screened Coulomb potential $V(r)=-r^{-1}e^{-C/r}$. The well known
Hellmann-Feynman theorem shows that all the bound states of the Coulomb
potential ($C=0$) remain bounded as $C$ increases. We derive a simple
approximate analytical expression for the eigenvalues for sufficiently small
values of the screening parameter $C$ and an approximate asymptotic
expression for the asymptotic behaviour of the s-state eigenvalues when $%
C\rightarrow \infty $. Present results are expected to resolve the
discrepancy about the spectrum of the quantum-mechanical model just
mentioned.
\end{abstract}

\section{Introduction}

\label{sec:intro}

There has recently been some interest in the screened Coulomb potential $%
V(r)=-r^{-1}e^{-C/r}$, $C>0$. Stachura and Hancock\cite{SH21} applied some
approximate methods to the radial Schr\"{o}dinger equation and conjectured
that there are critical values $C_{n}$ of the screening parameter $C$ so
that bound states disappear into the continuum when $C>C_{n}$. Later Xu et al%
\cite{MANY23} carried out accurate calculations by means of the powerful
generalized pseudospectral method and argued that there may not be such
critical values of the screening parameter. More precisely, the results of
Xu et al strongly suggest that all the bound states of the Coulomb potential
($C=0$) remain bounded for all values of $C>0$.

The purpose of this letter is the analysis of those contradictory results
and conclusions. In section~\ref{sec:Theoretical} we show that the well
known Hellmann-Feynman theorem (HFT)\cite{G32,F39} provides useful
information about the spectrum of the model just mentioned. We also propose
a simple approximate analytical expression for the eigenvalues of the
Schr\"{o}dinger equation for small values of $C$ and an asymptotic
expression for large values of this screening parameter. Finally, in section~%
\ref{sec:conclusions} we summarize our main results and draw conclusions.

\section{Theoretical analysis of the eigenvalue equation}

\label{sec:Theoretical}

In what follows, we focus on the radial part of the Schr\"{o}dinger equation
\begin{eqnarray}
&&-\frac{1}{2}\frac{d^{2}}{dr^{2}}\psi (r)+U(r)\psi (r)=E\psi (r),\;
\nonumber \\
&&U(r)=\frac{l(l+1)}{2r^{2}}+V(r),\;V(r)=-\frac{e^{-C/r}}{r},
\label{eq:Schrodinger_radial}
\end{eqnarray}
where $l=0,1,\ldots $ is the angular momentum quantum number and $C>0$. The
boundary conditions are
\begin{equation}
\lim\limits_{r\rightarrow 0}\psi (r)=0,\;\lim\limits_{r\rightarrow \infty
}\psi (r)=0.  \label{eq:boundary_cond}
\end{equation}

Stachura and Hancock\cite{SH21} stated that $U(r)$ becomes repulsive for a
sufficiently large value of $l$ and illustrated this fact for $C=0.1$ and $%
l=7$ in their figure 8. However, they did not appear to realize that
\begin{equation}
\lim\limits_{r\rightarrow \infty }rU(r)=-1.  \label{eq:lim_rU(r)}
\end{equation}
Their figure does not reveal this fact because the scale is rather too
coarse. Since $U(r)$ does not become repulsive for any value of $l$ and $C$
we cannot state that there are critical values of $C$. On the other hand,
the numerical results of Xu et al\cite{MANY23} suggest that all the bound
states of the Coulomb problem remain bounded for all values of $C$.

The HFT\cite{G32,F39} states that all the eigenvalues increase with $C$
\begin{equation}
\frac{dE}{dC}=\left\langle \frac{e^{-C/r}}{r^{2}}\right\rangle >0,
\label{eq:HFT_1}
\end{equation}
but it does not mean that some some of them may become positive by
increasing the value of $C$.

By means of the change of variables $\rho =r/C$ the eigenvalue equation can
be rewritten as
\begin{equation}
-\frac{1}{2}\frac{d^{2}}{d\rho ^{2}}\varphi (\rho )+\left[ \frac{l(l+1)}{%
2\rho ^{2}}-C\frac{e^{-1/\rho }}{\rho }\right] \varphi (\rho )=C^{2}E\varphi
(\rho ).  \label{eq:Schro_scaled}
\end{equation}
The HFT
\begin{equation}
\frac{dC^{2}E}{dC}=-\left\langle \frac{e^{-1/\rho }}{\rho }\right\rangle <0,
\label{eq:HFT_2}
\end{equation}
clearly shows that $C^{2}E$ decreases with $C$. If we take into account that
\begin{equation}
\lim\limits_{C\rightarrow 0}C^{2}E=0,  \label{eq:C->0}
\end{equation}
then we can safely conclude that all the bound states of the Coulomb
potential remain bounded as $C$ increases.

Both Stachura and Hancock\cite{SH21} and Xu et al\cite{MANY23} resorted to
approximations to $V(r)$ of the form
\begin{equation}
V^{[K]}(r)=-\frac{1}{r}\sum_{j=0}^{K}\frac{1}{j!}\left( -\frac{C}{r}\right)
^{j}.  \label{eq:V^[K]}
\end{equation}
The case $K=1$ is of particular interest because the resulting eigenvalue
equation
\begin{equation}
-\frac{1}{2}\frac{d^{2}}{dr^{2}}\psi (r)+\left[ \frac{l(l+1)+2C}{2r^{2}}-%
\frac{1}{r}\right] \psi (r)=E\psi (r),  \label{eq:Schrodinger_approx}
\end{equation}
can be solved exactly. Its eigenvalues are given by
\begin{equation}
E_{nl}=-\frac{1}{2\left( n+L-l\right) ^{2}},\;L=-\frac{1}{2}+\sqrt{\left( l+%
\frac{1}{2}\right) ^{2}+2C},  \label{eq:E_nl_approx}
\end{equation}
where $n=1,2,\ldots $ is the principal quantum number. This expression
yields satisfactory results for sufficiently small values of $C$. In fact,
the approximate eigenvalues shown in table~\ref{tab:EAP} for $C=0.1$ agree
reasonably well with the accurate ones obtained by Xu et al\cite{MANY23}.
This simple approximation suggests that there are bound states for all
values of $l$ when $C$ is sufficiently small (in particular, we may point
out the case $l=7$, $C=0.1$ discussed by Stachura and Hancock\cite{SH21} in
their figure 8).

In order to obtain an approximate asymptotic expression for the eigenvalues
for large values of $C$ we expand $V(r)$ about its minimum at $r=C$
\begin{equation}
V(r)\approx -\frac{1}{eC}+\frac{1}{2eC^{3}}\left( r-C\right) ^{2}.
\label{eq:V(r)_harmonic}
\end{equation}
The radial Schr\"{o}dinger equation with this approximate potential is
exactly solvable for $l=0$. Consequently, the eigenvalues $E_{n0}$ behave
approximately as
\begin{equation}
E_{\nu }\approx -\frac{1}{eC}+\sqrt{\frac{1}{eC^{3}}}\left( \nu +\frac{1}{2}%
\right) ,\;\nu =n-1=0,1,\ldots ,  \label{eq:E_nu_approx}
\end{equation}
for sufficiently large values of $C$. It is clear that
\begin{equation}
\lim\limits_{C\rightarrow \infty }CE_{\nu }=-\frac{1}{e},
\label{eq:CE_nu,C->0}
\end{equation}
in agreement with the result conjectured by Xu et al\cite{MANY23} from their
accurate numerical eigenvalues. The accuracy of the eigenvalues $E_{n0}$
under the harmonic approximation (\ref{eq:E_nu_approx}) increases with $C$
and decreases with $n=\nu +1$.

The eigenvalues given by the asymptotic harmonic approximation (\ref
{eq:E_nu_approx}) shown in table~\ref{tab:EAS} agree quite well with the
accurate results of Xu et al\cite{MANY23}. As stated above, the accuracy of $%
E_{\nu }$ increases with $C$ and decreases with $\nu $.

\section{Conclusions}

\label{sec:conclusions}

We have shown that the arguments put forward by Stachura and Hancok\cite
{SH21} about the existence of critical values of the screening parameter $C$
are not correct. The HFT (\ref{eq:HFT_2}) and equation (\ref{eq:C->0})
clearly show that the bound states of the Coulomb problem remain bounded as $%
C$ increases. Present approximate analytical expression (\ref{eq:E_nl_approx}%
) confirms this fact for small values of $C$. Besides, our
approximate eigenvalues agree with the accurate ones obtained by
Xu et al\cite{MANY23}. The latter results already follow the HFT
(\ref{eq:HFT_2}). In addition to what has just been said, we have
put forward a simple proof for the asymptotic behaviour of the
eigenvalues at large values of $C$ conjectured by Xu et al from
their numerical results.

\begin{table}[tbp]
\caption{Approximate eigenvalues $E_{nl}$ for $C=0.1$ obtained from equation
(\ref{eq:E_nl_approx}) ($\nu=n-l-1$)}
\label{tab:EAP}
\begin{center}
\par
\begin{tabular}{lD{.}{.}{6}D{.}{.}{6}D{.}{.}{6}D{.}{.}{6}}

\hline $\nu$ &  \multicolumn{1}{c}{$l=0$}&
\multicolumn{1}{c}{$l=1$} & \multicolumn{1}{c}{$l=2$}&
\multicolumn{1}{c}{$l=3$}\\ \hline

 0 &  -0.365    &   -0.117   &   -0.0541  &   -0.0308  \\
 1 &  -0.106    &  -0.0532   &   -0.0306  &   -0.0198  \\
 2 &  -0.0497   &  -0.0303   &   -0.0197  &   -0.0138  \\
 3 &  -0.0287   &  -0.0195   &   -0.0137  &   -0.0101  \\
 4 &  -0.0187   &  -0.0136   &   -0.0101  &   -0.00776 \\
 5 &  -0.0131   &  -0.0100   &   -0.00774 &   -0.00613 \\
 6 &  -0.00972  &  -0.00769  &   -0.00612 &   -0.00497 \\
 7 &  -0.00749  &  -0.00608  &   -0.00496 &   -0.00411 \\
 8 &  -0.00595  &  -0.00493  &   -0.00410 &   -0.00346 \\
 9 &  -0.00483  &  -0.00408  &   -0.00345 &   -0.00295 \\

 \end{tabular}
\end{center}
\end{table}

\begin{table}[tbp]
\caption{Eigenvalues $E_\nu$ obtained from the harmonic approximation (\ref
{eq:E_nu_approx})}
\label{tab:EAS}
\begin{center}
\par
\begin{tabular}{lD{.}{.}{8}D{.}{.}{8}D{.}{.}{8}}

\hline $C$ &  \multicolumn{1}{c}{$\nu=0$}&
\multicolumn{1}{c}{$\nu=1$} & \multicolumn{1}{c}{$\nu=2$}\\ \hline

$10^2$  & -3.38(-3)      & -2.8(-3)        &  -2.2(-3)   \\
$10^3$  & -3.583(-4)     & -3.39(-4)       &  -3.20(-4)   \\
$10^4$  & -3.6485(-5) & -3.588(-5)   &  -3.527(-5)   \\
$10^5$  & -3.6692(-6) & -3.650(-6)   &  -3.631(-6)   \\

\end{tabular}
\end{center}
\end{table}


\begin{thebibliography}{9}
\bibitem{SH21}  E. Stachura and N. Hancock, J. Phys. Commun. 5 (2021) 065004.

\bibitem{MANY23}  L. Xu, Jiao, L. G., A. Liu, H. E. Montgomery Jr., Y. K.
Ho, and S. Fritzsche, Phys. Lett A 483 (2023) 129064.

\bibitem{G32}  P. G\"{u}ttinger, Z. Phys. 73 (1932) 169-184.

\bibitem{F39}  R. P. Feynman, Phys. Rev. 56 (1939) 340-343.
\end{thebibliography}
\end{document}